# Chirality Imprinting and Spin-texture Tunability in Conformally Coated 3D Magnetic Nanostructured Metamaterials

Alexander Roberts[1], Huixin Guo[2], Joseph Askey[1], Vani Lanka[1], Arjen van den Berg[1], Dirk Grundler[2,3] and Sam Ladak[1]

1. School of Physics & Astronomy, Cardiff University, U.K.
2. Institute of Materials, EPFL, CH-1015 Lausanne.
3. Institute of Electrical and Micro Engineering, EPFL, CH-1015 Lausanne.

## Abstract

Three-dimensional (3D) magnetic nanostructures offer unprecedented opportunities for engineering emergent spin textures, but controlling their configuration remains a central challenge. Here we show that conformally coated Ni Nanotubes arranged in a woodpile geometry with lattice spacings ranging from 800 to 1200 nm, realised by two-photon lithography and atomic layer deposition, exhibit a geometry-tuneable balance between chiral and axial states. Magnetic force microscopy on the top layer of the 3D woodpile reveals that few-layer systems exhibit a chiral contrast whilst increasing the number of stacked layers drives a transition to an axial configuration with the change in state populations depending strongly on lattice spacing. Micromagnetic simulations demonstrate that chirality is not intrinsic to isolated tubes but is imprinted by spin textures formed in the substrate sheet film, which couple into the 3D network. As the sheet-film influence diminishes with increasing layer number, dipolar interactions dominate and stabilise the axial state. This two-stage mechanism of chirality imprinting followed by increasingly dominant dipolar interactions, provides clear control parameters for tailoring spin-texture populations. Our results establish conformally coated woodpiles as a reconfigurable 3D ferromagnetic metamaterial platform that can be exploited for data storage, magnonics, and neuromorphic computing.

## 1 Introduction

Over the past decade, research into three-dimensional (3D) magnetic nanostructures has grown rapidly[1,2], driven by their ability to host complex topological spin textures arising from geometric confinement[3], curvature-induced energies[4,5,6,7] and interactions between closely spaced elements[8,9]. This progress has laid the foundation for the development of magnetic metamaterials, engineered assemblies of "building blocks" arranged in intricate 3D architectures[10]. By precisely tuning the geometry and spacing of these elements, one can tailor their interactions, enabling the realization of systems with rich and tuneable magnetic phase diagrams[11,12]. The highly tuneable, extensive state space and large connectivity of these systems, when integrated into ultra-dense networks, unlocks a broad spectrum of technological applications, including racetracks for domain wall-based data storage[13,14,15], systems that harness controlled spin-wave propagation[11,16], large state space for neuromorphic architectures[17] and devices which harness non-reciprocal spin-wave dispersions[18, 19, 4]. Here, 3D architectures created from ferromagnetic metals stand out as they allow for exploitation in both spintronics and magnonic applications[19, 20,21].

A well-studied example of a magnetic metamaterial is artificial spin-ice (ASI), which consists of single-domain magnetic nanowires arranged in frustrated geometries[22]. While these systems have been extensively explored in two-dimensions (2D)[10,23], providing deep insights into model systems in statistical physics, their extension into 3D architectures has only recently begun to attract significant attention. For example, recent work has shown the placement of magnetic nanowires into diamond-bond lattice geometries provides degeneracy in nearest neighbour interactions[24] and allows the realisation of experimental charge-ordered ground states[12], that are otherwise impossible in 2D systems. Such systems also display unique spin-wave spectra[11] which can be tuned by controlling the populations of vertex states in the system[25]. Buckyball structures have also been extensively investigated, with recent Monte Carlo simulations revealing an intricate ground-state ordering[8], while micromagnetic simulations highlight their potential for reconfigurable magnonic applications[26]. Recent studies have also yielded a novel pyramid-based 3D-ASI system which supports an emergent artificial square ice[27]. More generally, Volkov et al. has studied 3D magnetic wireframe systems and established a correlation between the total magnetic vorticity of a system and its Euler characteristic[28]. Such pioneering work, when viewed in the context of tuneable interactions on a 3D lattice, offers a powerful framework for controlling both local spin textures and emergent global ground states.

Two-photon lithography (TPL) is a direct-write technology based on non-linear optical absorption providing confined polymerisation to realise arbitrary 3D nanostructures within a polymer[29,30]. Subsequent processing can be utilised to functionalise the polymer scaffolds with magnetic materials[16]. Line-of-sight techniques such as evaporation[24] or sputtering[31] offer a simple and efficient means to coat a 3D structure in magnetic material and can produce nanowires of novel crescent shaped cross-section[32]. However, from a technological point of view it is desirable to consider deposition schemes which allow for atomic-scale control and conformal deposition across a nanostructure, eliminating thickness gradients and shadowing effects. Recent developments in atomic layer deposition (ALD) have allowed for the conformal coating of nanostructures with Ni-based ferromagnets with excellent magnetic properties[33,4,34,]. This process has been used to coat TPL-fabricated scaffolds yielding a network of magnetic nanotubes arranged in a woodpile geometry[28] allowing for the realization of high-density ferromagnetic 3D lattices. These systems have also recently been studied within the context of magnonics, where high frequency modes were measured on the surface, but not in the bulk of the structure[35].

In this work, we investigate a 3D nanostructured magnetic metamaterial composed of conformally coated magnetic nanotubes arranged in an fcc woodpile geometry. Using scanning probe microscopy, we directly visualise the local spin textures and demonstrate that both the nanotube magnetisation and the overall magnetic configuration of the lattice depends characteristically on the number of stacked layers and the lattice parameter. Complementary micromagnetic simulations reveal that coupling between the woodpile structure and the conformal sheet film plays a decisive role, imprinting chiral spin textures that drive the emergence of complex, reconfigurable magnetic states. We show that a nearly complete tuning from chiral to aligned top layer configurations can be realized experimentally varying the submicron lattice spacing of the woodpile lattice.

## 2 Results & Discussion

3D Ni woodpile geometries were fabricated using a combination of two-photon lithography (TPL) and atomic layer deposition (ALD) as demonstrated schematically in Fig. 1(a) and 1(b). The details of the additive manufacturing methodology are presented in the Methods section. These woodpile structures consist of alternating orthogonally oriented polymer rods conformally coated in Ni with thickness t = 30 nm, Fig. 1(c), yielding an overall ferromagnetic nanotube with lateral feature size $d_{xy}$=400 nm and axial feature size $d_z$=990 nm. The resulting array of stacked nanotubes, shown in Fig. 1(c), possesses a lateral period $a_{xy}$ and a vertical period $a_z$, with $a_z = \sqrt{2}a_{xy}$. Critically, we note that whilst ALD conformally coats the woodpile structure, it also coats any exposed substrate yielding a sheet film upon the surface. The geometry of TPL and ALD produced nanotubes differ to previous magnetic systems[36, 37] as the TPL voxel takes an ellipsoidal geometry[29] with an aspect ratio (major/minor axes) of approximately 3. Woodpile structures were fabricated with lattice parameter $a_{xy}$= 800 nm, 1000 nm and 1200 nm, with a varying number $N$ of nanotubes stacked in the $z$-axis ($3 \leq N \leq 14$). Example scanning electron microscope images with lattice spacings of 800 nm, 1000 nm and 1200 nm for $N$ = 3 are shown in Fig. 1 (d-f) and for $N$ = 5 in Fig. 1 (g-i). It is important to note that the structures are partially embedded within the substrate (Fig. 1c) due to the TPL fabrication step, which provides mechanical anchoring. In the $N$ = 3 system (Fig. 1d–f), the two lower nanotube layers are therefore partially buried, while the uppermost layer is fully exposed above the surface. For larger $N$, all additional layers are fully protruding.

### 2.1 Investigating the spin texture upon the woodpile surface

Magnetic force microscopy (MFM) was performed on woodpile structures with varying $N$ and $a_{xy}$. The samples are first subjected to an external magnetic field protocol to ensure consistent field history. First, a field of $\mu_0 H = 290$ mT is applied parallel to the long-axis of the topmost layer of nanotubes, followed by removal of the field. Next a field of $\mu_0 H = 290$ mT is applied orthogonal to the long-axis of the top layer, followed by field removal. Finally, a field of $\mu_0 H = 500$ mT is applied out-of-plane with respect to the substrate and removed. MFM is then performed on this final remanent state. Figure 2 (a) shows example MFM contrast obtained for the woodpile with $a_{xy}$= 1000 nm and $N$ = 3 layers whilst Fig. 2 (b) shows a woodpile with $a_{xy}$= 1000 nm and $N$ = 14 layers. Supplementary Fig. 1. shows the same lattices measured with an inverted tip magnetisation, demonstrating the contrast is magnetic and arises from the stray fields of the nanotubes, as opposed to electrostatic forces or steep topographical gradients. The two samples exhibit markedly different contrast. The $N$ = 3 structure shows pronounced phase shifts of opposite sign on either side of the uppermost tube (zoomed image shown in Fig. 2c), consistent with a chiral state previously reported in literature[38]. In contrast, the $N$ = 14 structure displays regions of constant phase across the nanotube width, accompanied by bright and dark lobes along the tube axes at intersections with the second layer (zoomed image shown in Fig. 2d), characteristic of a more axial, single-domain configuration[38].

The individual contrast of each nanotube shown in Fig. 2 (a) largely possess the same chirality across the entire structure of $N$ = 3. The chiral state provides an efficient means to minimise the magnetostatic energy between adjacent tubes. The axial textures shown in Fig. 2 (b) appear to repeat, in a head-to-tail pattern throughout the woodpile of $N$ = 14, with small intermittent regions of chiral state remaining towards the array centre. Overall, this initial experiment reveals two distinct states, with their relative populations depending upon the number of nanotube layers $N$ stacked along the $z$-axis of the woodpile.

To investigate the nature of the transition from chiral to axial configurations more thoroughly, MFM was carried out upon a range of woodpile structures with $N$ = 3, 6, 10 and 14 and $a_{xy}$ = 800 nm, 1000 nm and 1200 nm. Figure 3(a-c) presents state maps, derived from MFM measurements, illustrating how the top-layer spin texture evolves with increasing $N$ and with variation in the lattice parameter $a_{xy}$. The raw MFM data are shown in Supplementary Fig. 2.

Across all values *of* $a_{xy}$, the $N$ = 3 systems exhibit a substantial preference for the chiral configuration observed (97.2% for $a_{xy}$ = 1200 nm, 96.9% for $a_{xy}$ = 1000 nm and 89.9% for $a_{xy}$ = 800 nm), with only few axial states being observed mostly at the tube ends. *As N* increases, the proportion of chiral states on the surface reduces in favour of the axial states. The extent to which axial states populate the top layer is found to depend critically upon $a_{xy}$. For the smallest investigated lattice parameter ($a_{xy}$ = 800 nm), changing from $N$ = 3 to $N$ = 6 layers yields an abrupt transition to a configuration whereby 60% axial state is observed, after which there is a more gradual increase in axial states with $N$, until the entire surface is populated with axial states at $N$ = 14. In contrast, for the largest lattice parameter ($a_{xy}$ = 1200 nm) the transition from chiral to axial state takes place across a larger range of $N$, reaching only 40% axial states at $N$ = 14. The state statistics are summarised in in Fig. 3(d). Overall, these results demonstrate the direct control of surface spin texture population through variation *of N* and $a_{xy}$.

### 2.2 Imprinting of Chirality Revealed by Micromagnetic Simulations

To gain insight into the origin of the experimentally observed spin textures, finite-difference micromagnetic simulations were performed using Mumax3[39]. We start with the simplest underlying unit, a single isolated nanotube with an elliptical cross-section, relaxed at random, to determine the minimum energy configuration in the absence of interactions with neighboring nanotubes. Supplementary Fig. 3 shows that isolated nanotubes, with dimensions matching those of the experimental woodpiles, stabilise multiple vortices along their vertical sidewalls. The large aspect ratio provides extended regions of reduced curvature, which favour vortex formation. From a top-down view (Supplementary Fig. 3 a,b), the resulting spin texture appears largely axial along the wire length, with occasional transitions in direction caused by interactions between counter-rotating vortices on opposite sidewalls (Supplementary Fig 3. c,d). The simulated axial state is not expected to produce the strong chiral contrast observed experimentally at low $N$, indicating that additional coupling mechanisms within the full woodpile geometry are important.

Having established the behaviour of isolated nanotubes, we next simulated geometries that more closely reproduced the experimentally realised woodpiles. Each simulation comprised a single unit cell in the *x*- and y-directions with periodic boundary conditions (3 repetitions in the x and y axes), corresponding to three in-plane repeated units. The geometries also considered the incomplete bottom-most unit cells that reflected the partially "buried" nanotubes of the experimental systems. As the ALD process led to a conformal deposition and introduced regions of sheet film between nanotubes, the geometries were simulated both with and without a Ni sheet film with $t$ = 30 nm, to understand the impact this had on the resulting spin textures. Due to memory constraints, simulations were restricted to systems with $N$ = 3 and $N$ = 4, which nevertheless captured some key trends observed experimentally. In all cases, the experimental field history was replicated to ensure direct comparison.

In the simulation, the structure with $N$ = 3 and $a_{xy}$ = 800 nm without sheet film (Fig. 4a) relaxes to an almost uniform axial state in the upper layer, with MFM contrast confined to the tube extremities and therefore failing to reproduce the strongly antisymmetric surface contrast of Fig. 2a. The transverse nanotube segments below the top layer (intermediate tubes), however, develop a weak chiral signature attributed to the stray field of the lowest, partially buried tubes, which remain uniformly magnetised. When a conformal sheet film is included (Fig. 4b), the upper tube instead stabilises a strongly chiral state, with positive $M_z$ on one side and negative $M_z$ on the other, with a simulated MFM signal that closely matches the experimental contrast. These comparisons suggest that at low lattice spacing, the sheet film introduces additional exchange and magnetostatic contributions at the tube/film interface, which resculpt the local energy landscape and favour a chiral state in the upper tubes. For $N$ = 4 (Fig. 4c), where the substrate sheet film reduced and further apart from the top layer tubes, the surface reverts to predominantly uniform axial states and the MFM contrast consists mainly of lobes at the tube intersections, in agreement with experiments performed on large *N* structures.

We now consider the *N* = 3, $a_{xy}$ = 1000 nm system. In the absence of a sheet film (Fig. 4d, left), the upper tubes relaxed to a predominantly axial state with a slight canting toward negative *x*-direction, arising from its exchange coupling to the intermediate tubes. This modest canting generates only weak MFM contrast (Fig. 4d, right), consistent with a non-chiral configuration. When a sheet film is introduced (Fig. 4e, left), the wider film regions now relax into vortices of like chirality. The magnetisation along the edges of these vortices couple directly into the partially buried tubes. As all film vortices share the same chirality, the induced tangential components enforce opposite signs of $M_y$ at the tube edges, producing a clear rotation of the magnetisation around the tube circumference and yielding ± $M_z$ contrast characteristics of a chiral state. Although the out-of-plane amplitude is somewhat smaller than in the 800 nm case, the MFM contrast ((Fig. 4e, right) remains qualitatively similar. For *N* = 4 (Fig. 4f), the sheet film lies farther beneath the surface, and the upper tube relaxes to a near-uniform axial state, giving an MFM signal consistent with corresponding large-*N* experiment.

Finally, we consider the *N* = 3 system with $a_{xy}$ = 1200 nm, with no sheet film. Here the upper tube appears to have opposite $M_y$ and $M_z$ components on each side of the tube (Fig 4g, left). Addition of the sheet film (Fig 4h), yields a uniform upper tube with canting and the *N* = 4 system (Fig 4i) yields a uniform axial upper state. Overall, the simulations provide qualitative agreement with experiment for low and intermediate lattice spacing, while at $a_{xy}$ = 1200 nm the reduced coupling and idealised geometry of the simulated unit cell limit direct comparison.

With the simulations clarifying the mechanism for chiral imprinting at small *N*, we now attempt to explain the global experimental trends observed in Fig. 3. To compare the stability of the two possible spin textures, the axial and chiral states, we define the total energy difference as $\Delta E(N, a_{xy}) = E_{axial} - E_{chiral}$ , where negative values indicate that the axial configuration is favoured, while positive values indicate a preference for the chiral configuration. For an isolated nanotube, micromagnetic simulations show that $\Delta E < 0$, and that the axial state is intrinsically lower in energy. For woodpiles with low values of *N*, the sheet film regions alter the energy balance by stabilising chiral configurations. We therefore introduce a film-mediated contribution that lowers the effective energy of the chiral configuration. As additional nanotube layers are added, an increasing fraction of the substrate is occupied by tube supports and buried sections, reducing the area of exposed substrate that can be coated as sheet film. If each added layer blocks a roughly constant fraction $p$ of the remaining accessible sheet-film area, then $A_N = A_0(1-p)^N = A_0 exp\,(-N/N_f)$ with $N_f = -1/\ln{(1-p)}$. With this in mind, we model the film mediated modification to energy, phenomenologically as $\Delta E_{film}(N, a_{xy}) = \Delta E^0_{film}(a_{xy}) exp\,(-N/N_f)$ where $\Delta E^0_{film}$ reflects the initial sheet film area (larger for wider lattice spacing $a_{xy}$) and $N_f$ is a characteristic decay constant defined above. This description captures the geometric reduction of accessible sheet-film area with increasing layer number, rather than a decay of magnetic coupling.

As the stack grows the presence and influence of the sheet film becomes progressively reduced and so magnetostatic interactions between neighbouring nanotubes accumulate and favour axial alignment. This is because a chiral surface state introduces significant out-of-plane magnetisation that generates both surface charges on the outer tube wall and volume charges at the interface with the underlying tube. In contrast, an axial state keeps $M_z$ = 0 upon the surface, and has flux closure on sidewalls, therefore minimising these sources of magnetostatic energy.

The magnetostatic interaction energy between magnetic moments decay as $1/r^3$ , so we write this contribution as $\Delta E_{dip}(N,a_{xy}) = -D\phi(N)a_{xy}^{-3}$, where $D$ is an effective coupling constant and $\phi(N)$ is a dimensionless function that captures how the net magnetostatic field from the underlying tubes increases with stack height $N$. Hence, we can write the total energy difference as:

$$\Delta E\left(N, a_{xy}\right) = \Delta E_{\text{single}} + \Delta E^{0}_{\text{film}}\left(\text{a}_{\text{xy}}\right) \exp\left(-N/N_f\right) - D\phi(N)\ a_{xy}^{-3}, \quad (1)$$

where $\Delta E_{single}$ is the intrinsic energy difference between axial and chiral states for a single, isolated nanotube.

The transition from the chiral to axial state occurs when $\Delta E(N,a_{xy}) = 0$. At low $N$, the positive sheet film term outweighs the single-nanotube offset and dipolar term yielding $\Delta E > 0$ and stabilising the chiral configurations. As $N$ increases, the film term decays, whilst the dipolar term grows, driving $\Delta E$ more negative and stabilising the axial state.

In the limit that the sheet-film contribution is negligible, the system cannot return to a regime where $\Delta E > 0$, since both remaining terms, the intrinsic single-nanotube offset and the dipolar term, favour the axial state. The transition therefore always occurs at finite $N$, when the decaying film contribution has just fallen below the combined axial-favouring terms. We denote this critical layer number by $N_c$. The condition for the crossover can be written as

$$\Delta E_{single} + \Delta E^{0}_{film}\left(a_{xy}\right) exp\left(-N_c/N_f\right) - D\phi(N_c)\ a_{xy}^{-3} = 0 \quad (2)$$

Thus, as the in-plane lattice spacing $a_{xy}$ increases, two physical effects manifest. Firstly, the initial sheet-film contribution is larger (since there is a larger available area for imprinting) and secondly, the dipolar contribution is weaker ($\propto a_{xy}^{-3}$). Consequently, a larger value of $N$ is required before the axial state dominates, so $N_c$ increases with $a_{xy}$. Overall, this toy model explains why the chiral state is rapidly lost for $a_{xy}$ beyond 800 nm, where $N_c$ is small but persists up to the largest measured value of $N$ for $a_{xy}$ = 1200 nm, where $N_c$ is larger. We note that this description is intended to explain the scaling behaviour rather than the detailed spin textures at the largest lattice spacing.

Taken together, our results demonstrate that conformally coated woodpiles offer multiple, geometry-based control parameters for engineering zero-field spin textures in 3D. Chiral configurations can be seeded via sheet-film imprinting at low $N$, while the subsequent suppression of chiral states is governed by dipolar interactions that are tuned by both the lattice spacing and number of stacked nanotubes in the $z$-axis. This combination of sheet-film imprinting, inter-tube spacing, and vertical layer number provides a powerful set of geometrically tuneable parameters for tailoring the balance between chiral and axial states. Such controllable spin-texture populations define these architectures as magnetic metamaterials, with direct implications for reconfigurable magnonic devices and for reservoir computing schemes where geometry-driven complexity can be harnessed for information processing.

## 3 Conclusion

In conclusion, we have demonstrated that 3D woodpile architectures conformally coated by a conventional metallic ferromagnet host two competing spin textures, chiral imprinted states and axially aligned states whose populations can be tuned through geometric parameters. For fewer stacked nanotubes in the $z$-axis, vortices in the conformal sheet film imprint chirality into the nanotubes, while this imprinting is progressively suppressed by dipolar interactions that favour axial alignment as the number of stacked nanotubes in the $z$-axis is increased. The crossover between these regimes depends strongly on the in-plane lattice spacing, providing a clear means for controlling spin-texture populations in three dimensions. These findings establish conformally coated woodpiles as a new class of magnetic metamaterial in which remanent spin textures can be reconfigured by geometry alone. Looking forward, such control opens pathways in information processing to reconfigurable magnonic devices, where chiral and axial states tailor effective fields and band structures of spin waves, and to spintronics devices in which the geometry-driven magnetic complexity controls spin currents for neuromorphic and reservoir computing.

## 4 Experimental Section

### Sample fabrication

The 3D polymer scaffolds were fabricated using a Photonic Professional GT+ (Nanoscribe GmbH) in CMi (EPFL). A negative photoresist, IP-Dip, with a refractive index of $n \approx 1.512$ at $\lambda = 780$ nm and room temperature, was used for two-photon lithography (TPL). IP-Dip photoresist was drop-cast onto the surface of a fused silica substrate with dimensions $25 \times 25 \times 0.7$ mm$^3$. The structures were written in a dip-in laser lithography (DiLL) configuration where a 63x, NA = 1.4 objective lens (Zeiss, Plan-Aprochromat) is brought into direct contact with the index-matched photoresist. A femtosecond-pulsed laser with free-space wavelength $\lambda = 780$ nm and an exposure power of 15 mW was used to realise the polymer woodpile lattice scaffolds. The beam was scanned across the sample using galvanometric mirrors. After writing, the structure was fully submerged in propylene glycol ether monomethyl acetate (PGMEA) for 20 minutes and isopropanol alcohol (IPA) for another 5 minutes. The samples were then allowed to dry in ambient conditions.

The ALD of Ni onto the polymer scaffolds follows the method outlined in[35]. A hot wall ALD system (Beneq TFS200) was used for this process. The 3D polymer scaffolds were positioned in the centre of the ALD chamber before being coated with a 5 nm thick layer of $Al_2O_3$ followed by a 30 nm thick layer of Ni.

### Scanning Electron Microscopy

The woodpile dimensions and burial within the sheet film was characterised using scanning electron microscopy (Hitachi SU8320 High Resolution scanning electron microscope), with all image analysis being performed using ImageJ[40].

### Magnetic Force Microscopy (MFM)

MFM data was captured using a scanning probe microscope (Bruker Dimension Icon) operated in tapping mode. Low moment tips (Bruker MESP-V2-LM) were magnetised along the tip axis using a 0.5 T permanent magnet. The image size was selected to capture the whole of the top surface of nanotubes in the woodpiles so varied based on the value of $a_{xy}$, with the smallest scans having dimensions of 8 µm x 8 µm and the largest 16 µm x 16 µm. Owing to the 3D topography of the woodpile structure, the scanning and feedback parameters were carefully optimised during the AFM scan for each image. The lift height for the second pass was set at 70 nm for all scans to ensure sufficient image resolution without introducing artefacts which result from the tip striking the sample. Separate scans with a reversed tip magnetisation were performed to verify consistency of the contrast and control for artefacts, as scans with inverted tip magnetisation are expected to yield MFM images where the sign of the phase shifts is inverted, which confirms the magnetic origin of the contrast.

## Micromagnetic Simulations

Micromagnetic simulations were performed using Mumax3[39], a GPU-accelerated package, which uses a finite-difference discretization of the simulation space to solve the Landau-Lifshitz-Gilbert (LLG) equation,

$$\frac{d\mathbf{M}}{dt} = -\gamma_0(\mathbf{M} \times \boldsymbol{H}_{\text{eff}}) + \alpha(\mathbf{M} \times \frac{d\mathbf{M}}{dt}). \quad (3)$$

Where **M** represents the normalised magnetisation vector, $\gamma_0$ = 2.211 x $10^5$ $mA^{-1}s^{-1}$ is the gyromagnetic ratio, α is the phenomenological damping constant, and $\boldsymbol{H}_{\text{eff}}$ is the effective field,

$$\boldsymbol{H}_{\text{eff}} = \frac{-1}{\mu_0 M_s}\frac{\delta_{\mathbf{M}}[\boldsymbol{\varepsilon}]}{\delta\mathbf{M}}, \quad (4)$$

with ε representing the energy density, $\frac{\delta_{\mathbf{M}}[\boldsymbol{\varepsilon}]}{\delta\mathbf{M}}$ is the variational derivative of the energy density with respect to the normalised magnetisation, $Ms$ is the saturation magnetisation, and $\mu_0$ = 1.2566 x$10^{-6}$ $Hm^{-1}$. The material parameters were chosen to correspond with bulk Ni: $M_s$ = 490 $kAm^{-1}$ and exchange stiffness, $A_{ex}$ = 8 $pJm^{-1}$. The damping constant was set to α = 1 for all relaxation simulations to reduce computational time without sacrificing accuracy. The exchange length of bulk Ni is calculated as:

$$l_{ex} = \sqrt{2A_{\text{ex}}/\mu_0 M_s^2}, \quad (5)$$

which is equal to ~7 nm for bulk Ni. Thus, the discretization cell size was set to 5 nm in all three axes to both satisfy the exchange length and form an appropriate approximation of the curved nanotube geometries, whilst ensuring reasonable computational time and not exceeding available video memory. Contributions arising from the magnetocrystalline anisotropy (MA) were assumed to be negligible due to the averaging of local MA axes in the polycrystalline Ni film. Given this consideration, the total energy density can then be written as the sum of exchange and magnetostatic energies:

$$\varepsilon = -A_{\text{ex}}\boldsymbol{M} \cdot \nabla^2\mathbf{M} - \frac{1}{2}\mu_0 M_s(\boldsymbol{H}_{\text{d}} \cdot \mathbf{M}), \quad (6)$$

where $\boldsymbol{H}_d$ is the demagnetising field.

The geometries shown were created using built-in elementary geometric shapes and logic operations within MuMax3, with nanotube dimensions given by SEM measurements.

## Magnetic Force Microscopy Simulations

With the probe magnetised and oscillating along the *z*-axis, MFM contrast is proportional to $\varphi \propto \partial_z^2 H_z$. The sample stray field at the tip's position ($\boldsymbol{r}_t$) is given by

$$\mathbf{H}(\mathbf{r_t}) = \sum_i \frac{3(\boldsymbol{\mu}_i \cdot \mathbf{r})\mathbf{r}}{r^5} - \frac{\boldsymbol{\mu}_i}{r^3} \quad (7)$$

Where $\boldsymbol{\mu}_i$ is the normalised $i^{th}$ moment in the simulation object at position $\mathbf{r}_j$, and **r** is given by $\mathbf{r} = \mathbf{r}_t - \mathbf{r}_j$.

The location of the probe apex for $\ell = 0$ nm is calculated numerically by constructing a surface representing the MFM probe, and the upper surface of the simulation object. A binary search algorithm determines the smallest offset from the substrate plane where the surfaces do not intersect, approximating the position of the tip's apex on the AFM pass of the measurement representing the simulated topography measurement.

The stray field is subsequently calculated at $\ell$ - 1 nm, $\ell$, and $\ell$ + 1 nm above the simulated topography measurement, and the MFM contrast is then computed using:

$$\varphi = \partial_z^2 H_z = H_z^{\ell+1} - 2H_z^{\ell} + H_z^{\ell-1} \quad (8)$$

**Acknowledgements**

SL acknowledges funding from the EPSRC (EP/L006669/1, EP/R009147) and Leverhulme Trust (RPG-2021-139). DG acknowledges funding from SNSF( grant 197360). Experimental support by D. Bouvet and further staff members of the Centre of MicroNano Technology (CMi) at EPFL is gratefully acknowledged.

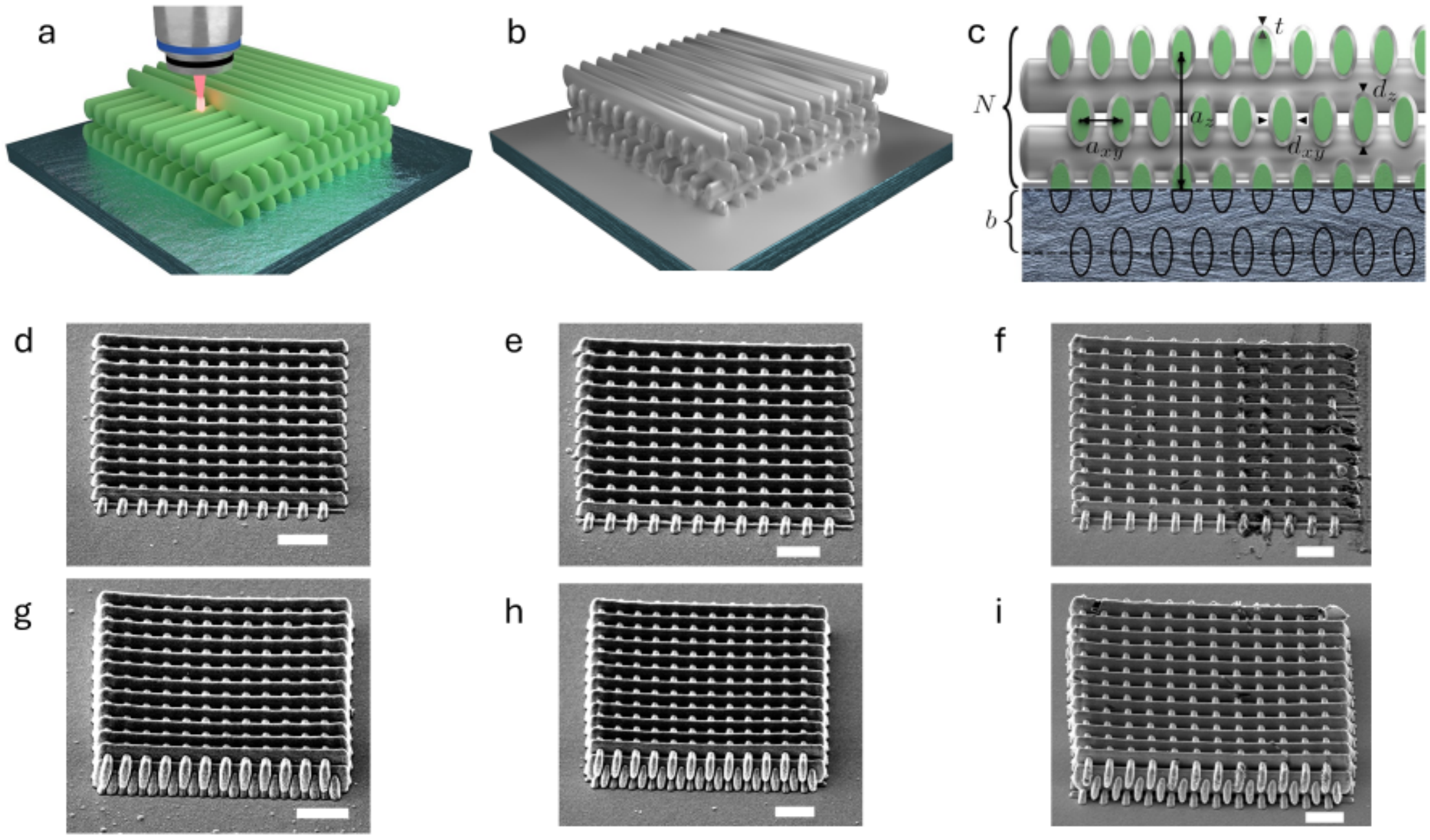

**Figure 1:** ***Fabrication and imaging of a 3D nanostructured magnetic metamaterial.*** *(a) Two-photon lithography is used to define a polymer (green) scaffold. (b) Atomic layer deposition (ALD) conformally coats the scaffold and substrate with Ni (grey). (c) Cross-sectional schematic of the 3D woodpile lattice with lateral and axial parameters $a_{xy}$ and $a_z = \sqrt{2}a_{xy}$ respectively, thickness t, number of stacked nanotubes in z-axis N, lateral and axial feature sizes $d_{xy}$ and $d_z$ and 'burial' of nanotubes within the substrate b. SEM images of woodpile with number of layers N = 3 and with lattice parameter (d) $a_{xy}$ of 800 nm, (e) 1000 nm and (f )1200 nm. SEM images of woodpile with an increased number of layers N = 6 and with lattice parameter (g) 800 nm, (h) 1000 nm and (i) 1200 nm. SEMs images were captured at a 45 degree angle. All scale bars represent 2 μm.*

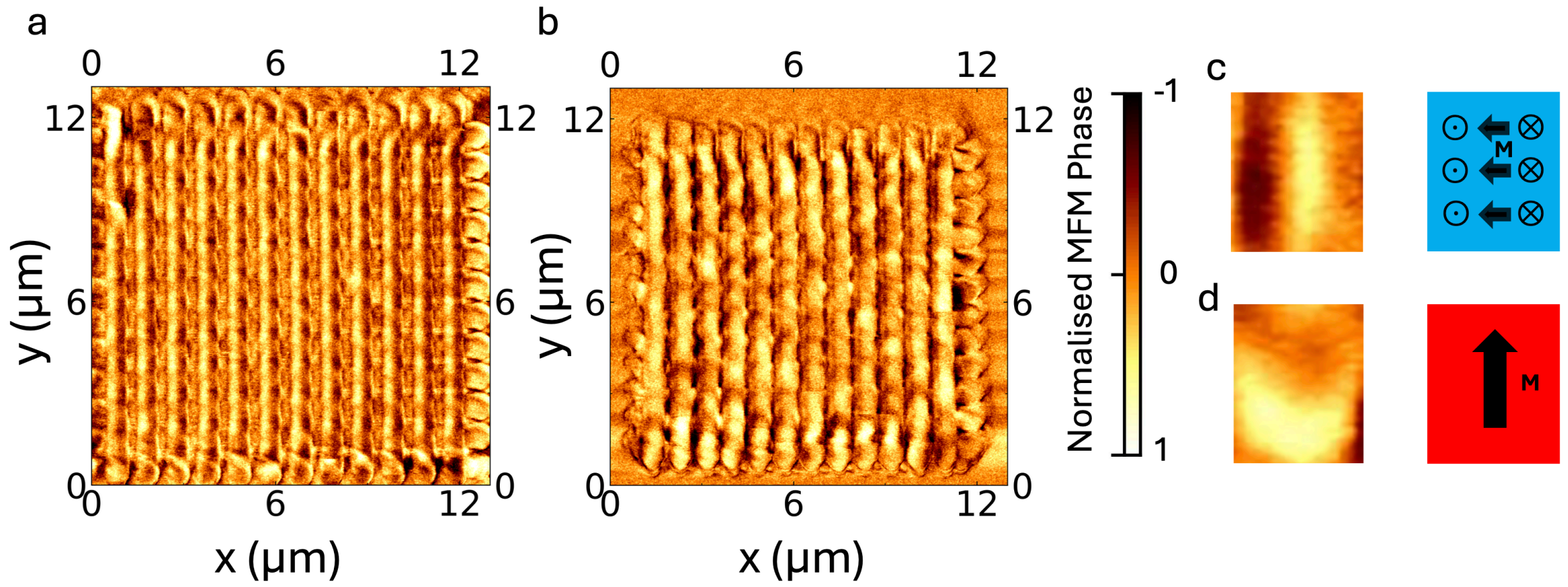

**Figure 2***: Magnetic force microscopy (MFM) of 3D nanostructured metamaterial.* *(a) MFM for system with $a_{xy}$ = 1000 nm N = 3. Opposing contrast is observed at the nanotube edges indicative of a chiral configuration. (b) MFM for system with $a_{xy}$* = 1000 nm, *N* = 14. *Two distinct states are observed; a chiral state, seen in (a), and a second state suggestive of axial contrast. (c) Zoomed contrast of a chiral state (left) with a schematic showing the suggested magnetisation configuration (right). (d) Zoomed contrast of the axial state (left) with a schematic showing the suggested magnetisation configuration on the tube top.*

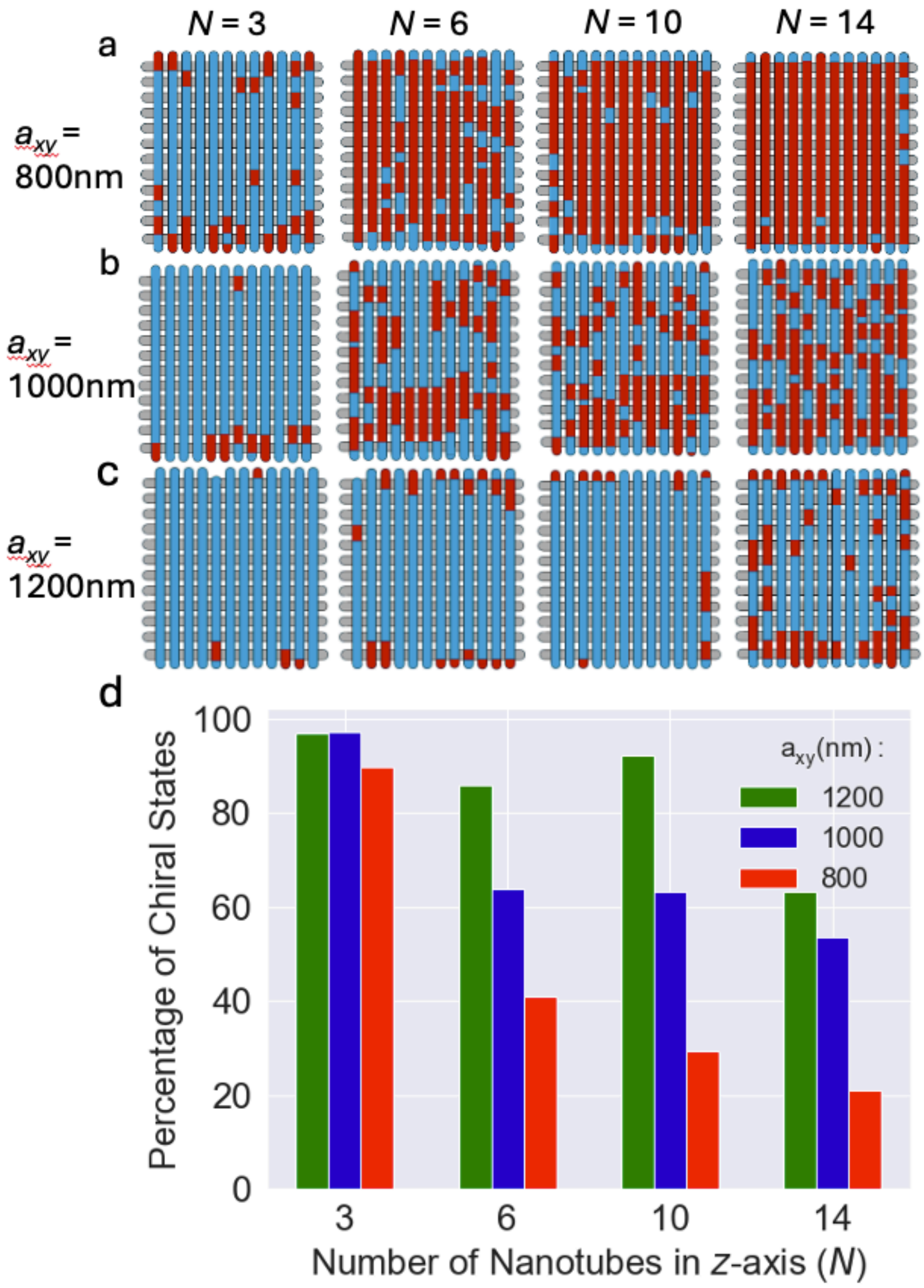

***Figure 3: Evolution of spin textures in 3D nanostructured metamaterial.*** *(a-c) Distribution of the two configurations across the topmost layer of the woodpile geometries, based upon MFM measurements, with the chiral configuration shown in blue and the axial configuration shown in red. (d) State statistics of the woodpile top-layer as function of lattice parameter* $a_{xy}$ *and number of layers N. In all cases the chiral state is dominant for low N, The rate at which the configuration changes from chiral to axial, depends critically upon the lattice spacing* $a_{xy}$*, with a more rapid cross-over occurring for low* $a_{xy}$*.*

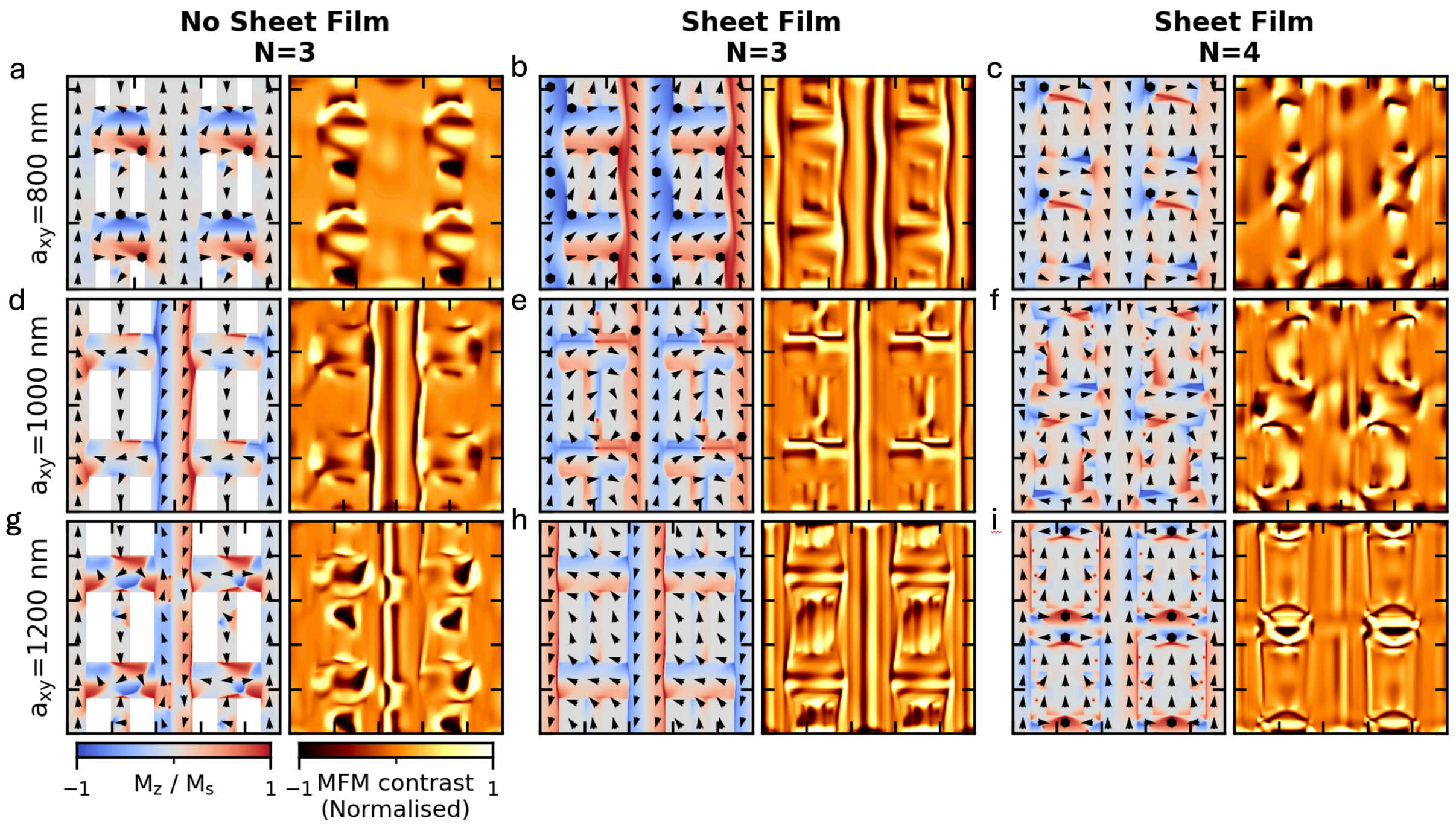

**Figure 4: Micromagnetic simulations and chiral imprinting.** Spin texture is shown in left panels and calculated magnetic force microscopy contrast is shown in right panels. (a) Simulation for *N = 3 and $a_{xy}$ = 800nm with no sheet film. Uppermost tube has axial state. (b)* Simulation for *N = 3 and $a_{xy}$ = 800nm with sheet film. The sheet film indirectly stabilises a chiral state in the upper tube. (c)* Simulation for *N = 4 and $a_{xy}$ = 800nm. Here though a small amount of sheet film remains, its impact is negligible and an axial state is stabilised on the upper tube. (d)* Simulation for *N = 3 and $a_{xy}$ = 1000nm with no sheet film. (e)* Simulation for *N = 3 and $a_{xy}$ = 1000nm with sheet film. Here vortices on the planar sheet film are found to imprint a chiral state into the uppermost tube. (f)* Simulation for *N = 4 and $a_{xy}$ = 1000nm. The impact of sheet film has diminished due to further distance and an axial state is stabilised. (g)* Simulation for *N = 3 and $a_{xy}$ = 1200nm with no sheet film. Uppermost tube has alternating $M_y$ components. (h)* Simulation for *N = 3 and $a_{xy}$ = 1200nm with sheet film. The upper surface shows an axial state with some canting. (i)* Simulation for *N = 4 and $a_{xy}$ = 1200nm. An axial state is stabilised in the top tube.*

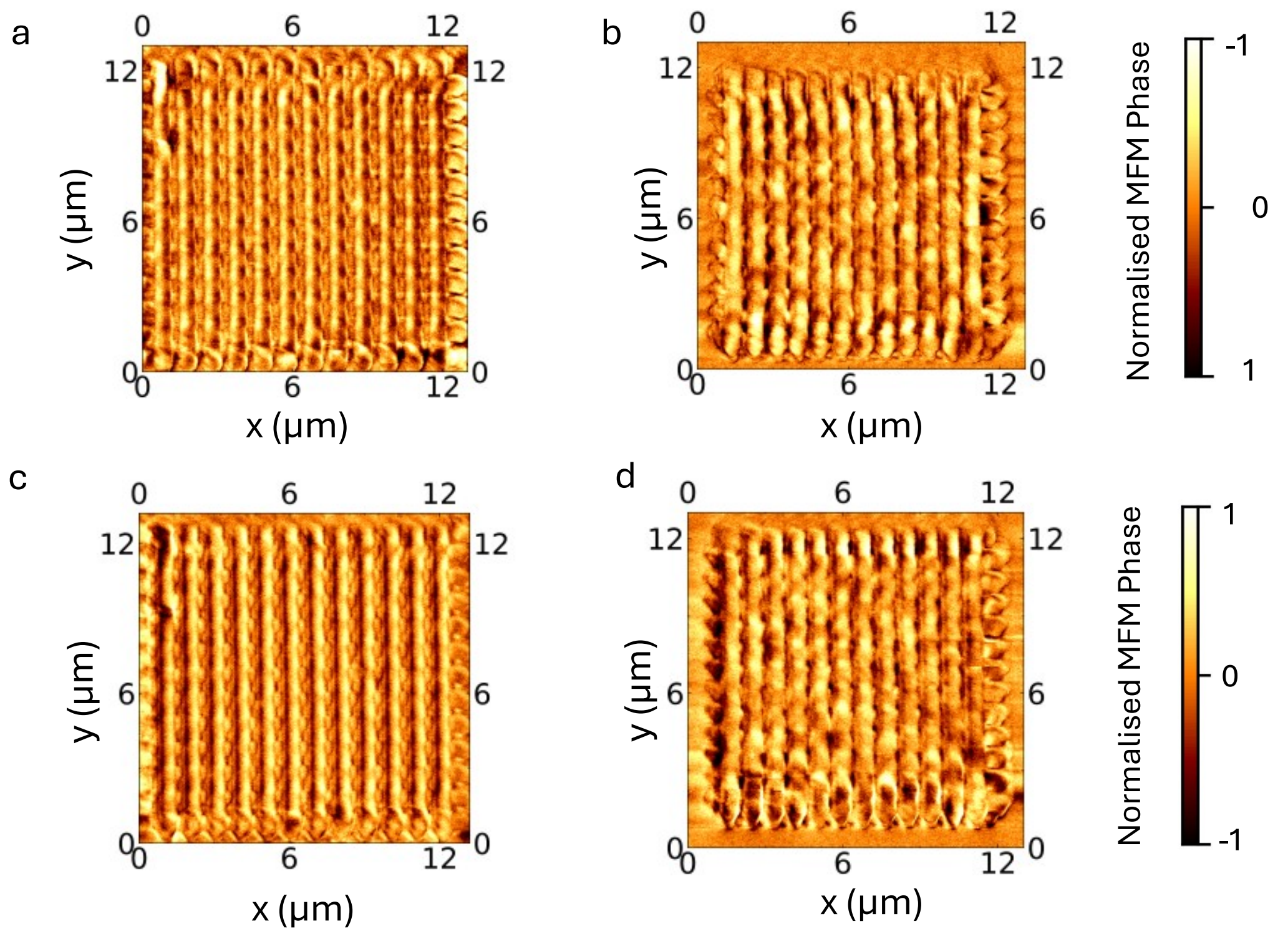

**Supplementary Figure 1:** *Magnetic force microscopy images taken with inverted tip magnetisations. (a) Woodpile with* $a_{xy}$ *= 1000 nm and N = 3 and (b) Woodpile with* $a_{xy}$ *= 1000 nm N = 14. (c) The same woodpile as (a) but with inverted tip magnetisation, and (d) the same woodpile as (b) but with inverted tip magnetisation. The sign of the MFM contrast along the sides of the topmost layer of nanotubes inverts. Both sets of scans demonstrate inversion of MFM contrast and therefore confirmation of the magnetic origin of the signal.*

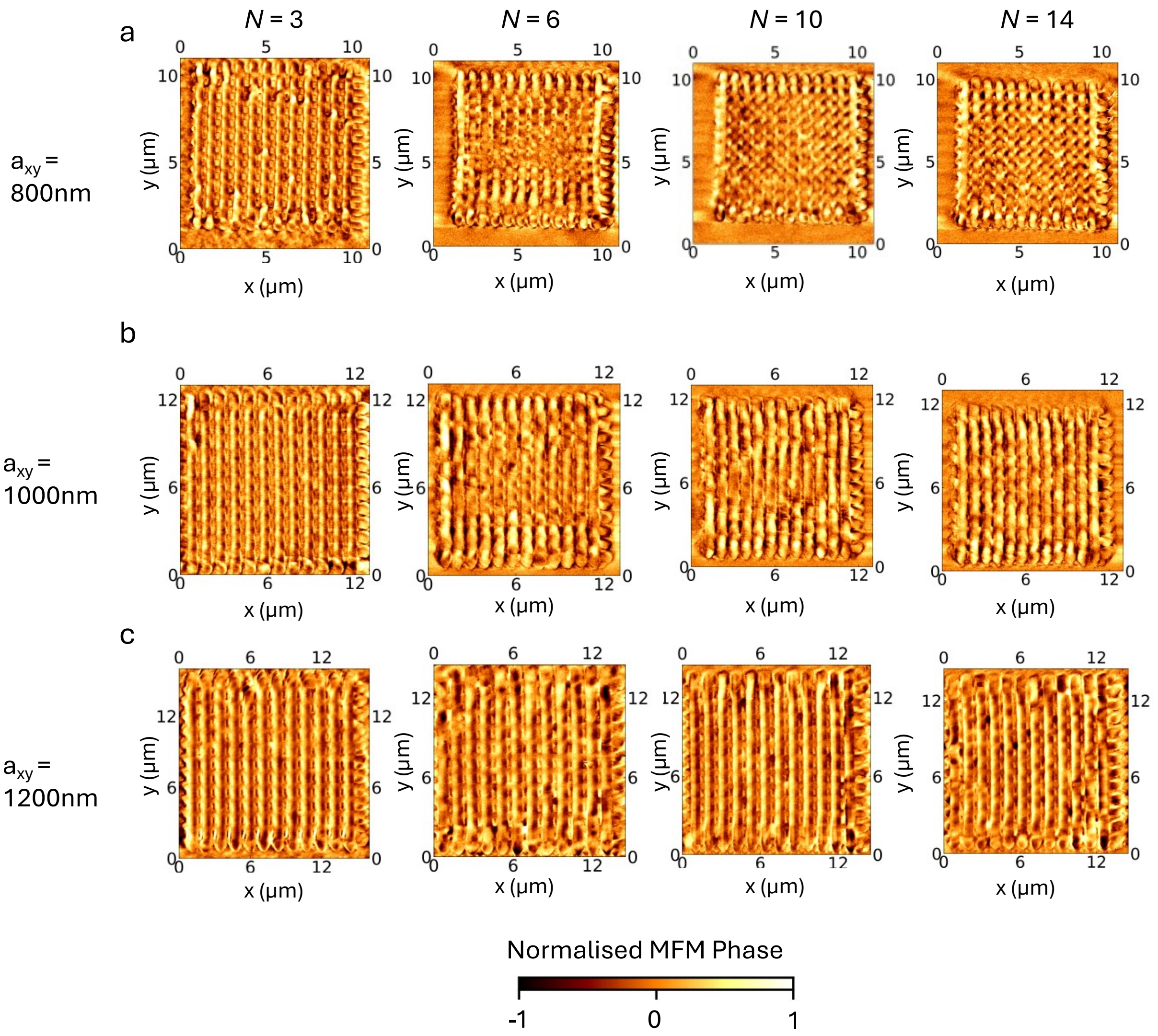

***Supplementary Figure 2:*** *(a-c) MFM images of the top-layer of woodpile nanotube structures with varying lattice parameter ($a_{xy}$) and number of nanotubes stacked in Z (N) used for creation of the schematics shown in Fig 3.*

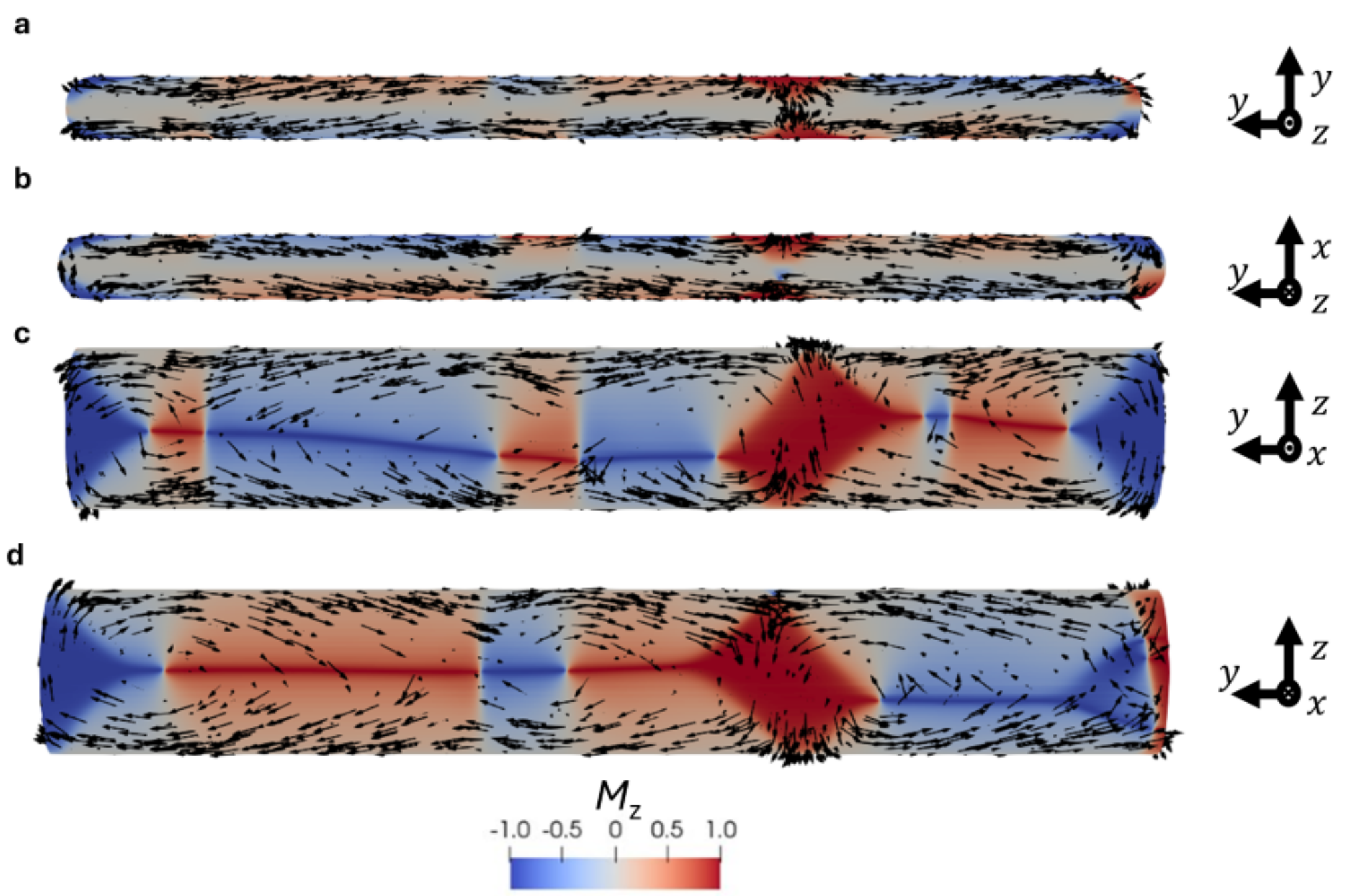

***Supplementary Figure 3:*** *Micromagnetic simulations of a single nanotube with an identical geometry to the nanotubes in the experimental woodpile system, and a length of 7 µm, showing different orientations of the same tube. (a) and (b) show the view of the top and bottom respectively, where it can be seen that the magnetization mostly orients along the tube in an axial manner, with a domain wall forming as a result of a magnetic vortex forming on the. (c,d) Sidewall configurations showing the formation of vortices.*